\documentclass[aps,prd,twocolumn,groupedaddress]{revtex4-1}

\usepackage{graphics,graphicx}
\usepackage{amsmath, amssymb}
\usepackage{natbib,hyperref}
\usepackage{mathtools}

\begin{document}

\title{Beyond Mean Field: Fluctuation Diagnostics and Fixed-Point Behavior}

\author{Pok Man Lo}
\affiliation{Institute of Theoretical Physics, University of Wroclaw,
PL-50204 Wroc\l aw, Poland}
\email{pokman.lo@uwr.edu.pl}

\begin{abstract}
We develop theoretical diagnostics for the breakdown of mean-field theory, 
demonstrate how spatial structure and finite interaction ranges enter the effective description, 
and show how these scales qualitatively modify the renormalization-group flow.
\end{abstract}

\maketitle

\section{Introduction}

Universality organizes critical phenomena through fixed points and scaling laws~\cite{Goldenfeld:1992qy,Cardy:1996xt}, but it does not specify how close a system must be to a critical point before universal behavior becomes physically relevant. This is controlled by the size of the fluctuation-dominated region, which is non-universal and sensitive to microscopic details. In many systems—including those relevant to dense QCD matter—this region can be narrow, so identifying a universality class alone does not guarantee that critical behavior will be observable.

A more practical diagnostic comes directly from the Ginzburg–Landau (GL) framework. Within GL theory, the validity of mean-field (MF) solutions can be assessed by comparing fluctuation contributions to the MF order parameter, known as the GL criterion. This provides a quantitative measure of where fluctuations overwhelm MF behavior, without relying on universality arguments. Importantly, the criterion also applies to crossover transitions, making it broadly useful for realistic systems.

Fluctuation estimates can be further refined by retaining gradient terms in the GL functional. Standard Landau treatments often assume spatial homogeneity by neglecting these terms, an approximation commonly justified by the associated increase in kinetic energy. However, this assumption fails once non-local interactions are present, or more generally when quantum corrections are included. In such cases, spatially varying MF configurations arise naturally. Spatial structure is therefore not an exotic feature, but an intrinsic part of a consistent GL description and must be included in any reliable analysis.

Microscopic interactions may also introduce intrinsic momentum scales, which influence renormalization-group (RG) flows and the locations of fixed points. Understanding how model-dependent structures shape RG trajectories helps identify which features of an interaction kernel reorganize the effective theory at intermediate scales.

This article is intended as pedagogical notes that develop these ideas in a simple and transparent way. We focus on practical methods to quantify the breakdown of mean-field theory, assess the role of spatial structure, and analyze how RG flows respond to microscopic inputs.

\section{Revisiting the Ginzburg-Landau criteria}

The GL criterion focuses on the ratio
\begin{align}
    R_{\rm GL}
    = \frac{\langle (\Delta \phi)^2 \rangle}{\bar{\phi}^2},
    \label{eq:GL01}
\end{align}
where $\phi$ is the order-parameter field and $\bar{\phi}$ its mean value.
When $R_{\rm GL} \gg 1$, fluctuations are no longer negligible and the MF
approximation fails. In standard textbook treatments~\cite{Goldenfeld:1992qy}
this ratio is used to estimate the critical dimension. Here we take a different
perspective: we use the criterion to identify the region in thermodynamic
variables where MF theory does not apply. This region serves as a proxy for the
extent of the fluctuation-dominated domain in the phase diagram.

We consider two ways to estimate $\langle(\Delta\phi)^2\rangle$: one in
configuration space and one in momentum space. In the first formulation, we use
the static susceptibility to quantify the fluctuations,
\begin{align}
    \begin{split}
    \chi_{\rm static}
        &= \beta \int d^3x\, G_c(\vec{x}) \\
        &= \beta\,\tilde{G}_c(\vec{k}=0),
    \end{split}
    \label{eq:sus}
\end{align}
where $G_c(\vec{x})=\langle \phi(\vec{x})\phi(\vec{0})\rangle_c$ is the connected
two-point function and $\tilde{G}_c(\vec{k})$ its Fourier transform.

The GL ratio can then be written as
\begin{align}
    R_{\rm GL}^{(1)}
    = \frac{\int d^3x\, G_c(\vec{x})}{\int d^3x\, \phi(\vec{x})^2}
    \;\to\;
    \frac{\beta^{-1}\chi_{\rm static}}{V_\phi\,\bar{\phi}^2},
    \label{eq:R1}
\end{align}
where $V_\phi$ denotes the region over which $\phi(\vec{x})$ can be approximated
by its mean value $\bar{\phi}$.~\footnote{In Ref.~\cite{Goldenfeld:1992qy} (Eq.~6.7), the choice $V_\phi=\xi^3$,
with $\xi$ the correlation length, is used in estimating the GL ratio. We
discuss alternative choices below.}
Importantly, $V_\phi$ should not be taken to infinity. 
In realistic situations the finite size of domains or the presence of gradient terms—unavoidable in the quantum case—limits the spatial extent over which a uniform field
approximation is valid.

Within a homogeneous MF theory, $\bar{\phi}$ is obtained by minimizing the mean-field
potential $U_{\rm MF}(\phi)$. The static susceptibility is then given by the
inverse curvature,
\begin{align}
    \chi_{\rm static}
    \;\to\; \left(U_{\rm MF}''\right)^{-1}
    = \left(
        \frac{\partial^2 U_{\rm MF}(\phi)}{\partial\phi\,\partial\phi}
      \right)^{-1},
\end{align}
understood to be evaluated at the MF solution $\phi=\bar{\phi}$.

For completeness, we include a minimal derivation. MF theory may be viewed as a
zero-dimensional field theory,
\begin{align}
    Z(h) = \int d\phi \, e^{-\beta V \left[ U(\phi)-h \phi \right]} \equiv
    e^{-\beta V f(h)}.
\end{align}
The mean field is computed from
\begin{align}
    \langle \phi \rangle
    = \frac{1}{\beta V} \frac{\partial \ln Z}{\partial h}
    = -\frac{\partial f}{\partial h}.
\end{align}
The static susceptibility is an intensive observable obtained from the second derivative of $f(h)$,
\begin{align}
    \chi_{\rm static}
    = -\frac{\partial^2 f}{\partial h \partial h}
    = \beta V \left( \langle \phi^2 \rangle - \langle \phi \rangle^2 \right).
\end{align}
An effective potential is formally defined via the Legendre transform of $f(h)$~\cite{Negele:1988vy},
which amounts to exchanging the roles of $h$ and $\phi$,
\begin{align}
    \gamma(\phi) = f(h) + h \phi,
\end{align}
where $h$ is understood as a function of $\phi$. The canonical relation is
\begin{align}
    h = \frac{\partial \gamma}{\partial \phi}.
\end{align}
Differentiating once more yields the inverse-curvature relation,
\begin{align}
    -\frac{\partial^2 f}{\partial h \partial h}\,
    \frac{\partial^2 \gamma}{\partial \phi \partial \phi} = 1.
\end{align}
The MF approximation corresponds to identifying $\gamma(\phi)=U_{\rm MF}(\phi)$, and therefore
\begin{align}
    \chi_{\rm static} = \left(U_{\rm MF}''\right)^{-1}.
\end{align}

\begin{figure}
    \centering
    \includegraphics[width=0.5\textwidth]{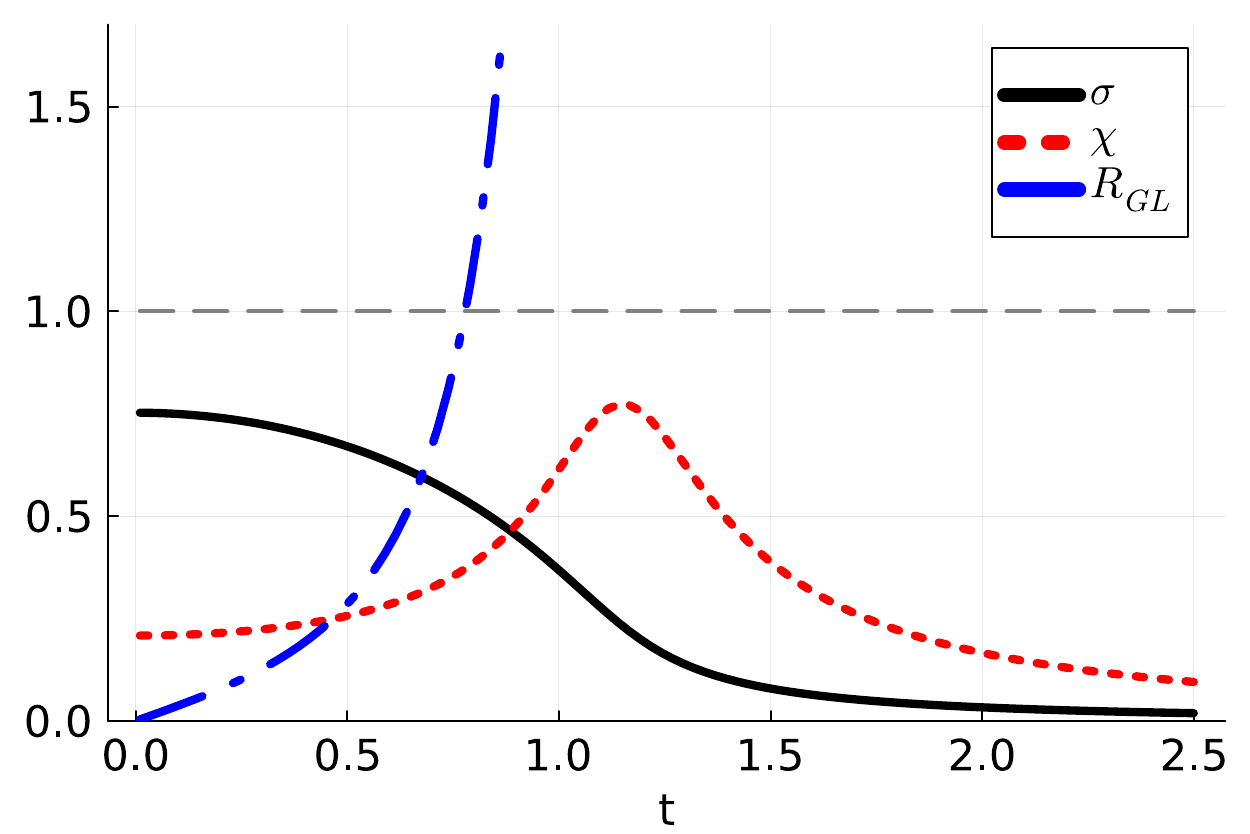}
    \includegraphics[width=0.5\textwidth]{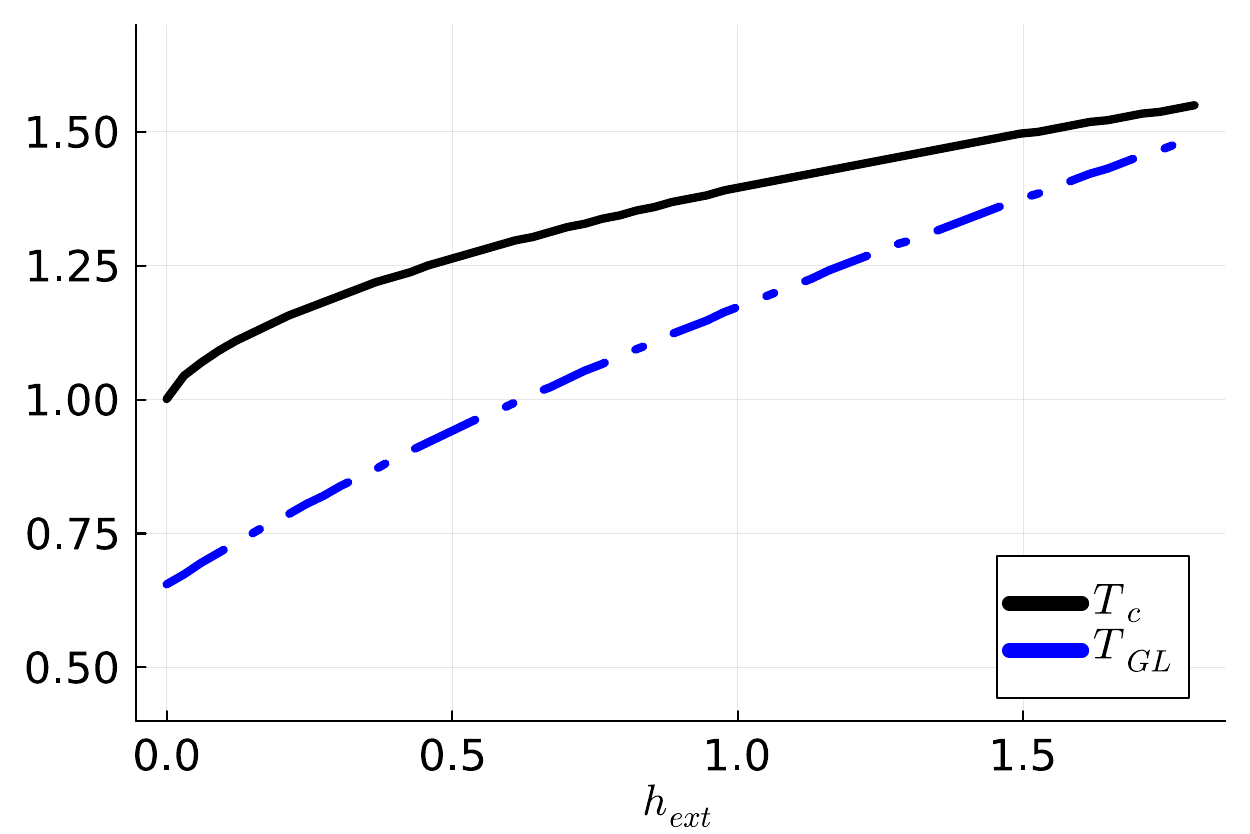}
    \caption{Demonstration of the GL ratio in a linear sigma model. Top:
    temperature dependence at fixed explicit breaking $h$. Bottom:
    pseudo-critical temperature as a function of $h$. The interval
    $\vert T_{\rm GL}-T_c \vert$ highlights the region where MF theory is not
    reliable.}
    \label{fig:fig1}
\end{figure}

We illustrate the robustness of the GL ratio by examining a simple linear
sigma model (LSM). The prototypical Landau potential is
\begin{align}
    U_{\rm LSM}(\sigma) = -(1 - t^2)\,\sigma^2 + \sigma^4 - h\,\sigma ,
\end{align}
with $t = T/T_0$. Here $T_0$ is the only intrinsic scale of the model and is
used as our unit of energy. The equation of motion can be solved numerically
for arbitrary external field $h$, and the resulting mean field, static
susceptibility, and GL ratio are shown in Fig.~\ref{fig:fig1}. In this example
we set $V_\phi \, T_0^3 = 1$.

As the temperature increases, the GL ratio exceeds unity at a temperature
$T_{\rm GL}$ that lies below the pseudo-critical temperature $T_c$, the latter
identified from the peak of the static susceptibility. We further track the
dependence of $T_{\rm GL}$ on the external field $h$ and compare it with
$T_c(h)$. The line of pseudo-critical temperatures plays the role of a phase
diagram, while the area $|T_{\rm GL} - T_c|$ quantifies the region where
MF theory ceases to be reliable. As expected, this region widens
when approaching the critical point, and the GL ratio provides a clean and
quantitative diagnostic.

We now turn to an alternative, and in fact, more general way of estimating fluctuations.
Consider the fluctuation of the field at a point:
\begin{align}
    \begin{split}
        G_c(\vec{x}\to \vec{0}) &= 
        \lim_{\vec{x}\to 0} \langle \phi(\vec{x})\,\phi(\vec{0}) \rangle_c \\
        &= \int \frac{d^3 k}{(2\pi)^3}\,\tilde{G}_c(\vec{k})
    \end{split}
\end{align}
where $\tilde{G}_c(\vec{k})$ is the Fourier transform of $G_c(\vec{x})$. 
Such fluctuation is dominated by UV momenta. To control this behavior, we define a
regulated version,
\begin{align}
    G_c^{\Lambda}(\vec{0}) =
    \int^{\Lambda} \frac{d^3 k}{(2\pi)^3}\,
    \tilde{G}_c(\vec{k}) ,
\end{align}
and construct the corresponding GL ratio
\begin{align}
    R_{\rm GL}^{(2)}(\Lambda)
    = \frac{\displaystyle \int^{\Lambda} \frac{d^3 k}{(2\pi)^3}\,
    \tilde{G}_c(\vec{k})}{\bar{\phi}^2} .
    \label{eq:R2}
\end{align}
In this form, $\Lambda$ is best interpreted as a physical scale specifying the
range of fluctuations to be included—not as a UV cutoff intended to be removed.
Misinterpreting it as such leads to unphysical conclusions.

At first sight, the definition above appears to be incompatible with the GL ratio in
Eq.~\eqref{eq:R1}. We now show explicitly that no inconsistency exists. For
illustration, we consider a simple Gaussian model with
\begin{align}
    \tilde{G}_c(\vec{k})
    = \frac{1}{\beta}\frac{1}{\vec{k}^2 + m^2} .
\end{align}
The regulated integral can be evaluated analytically:
\begin{align}
    \int^{\Lambda} \frac{d^3 k}{(2\pi)^3}\,\tilde{G}_c(\vec{k})
    = \frac{1}{2\pi^2}\frac{1}{\beta}
    \left( \Lambda - m \tan^{-1}\frac{\Lambda}{m} \right) .
    \label{eq:analytic01}
\end{align}
The GL ratio in Eq.~\eqref{eq:R1} probes long-wavelength fluctuations within a
finite volume $V$. To reproduce the same low‑momentum window, we choose
$\Lambda = \Lambda_{\rm IR}$ such that
\begin{align}
    \int^{\Lambda_{\rm IR}} \frac{d^3 k}{(2\pi)^3} = \frac{1}{V} ,
\end{align}
which gives
\begin{align}
    \Lambda_{\rm IR}^3 = \frac{6\pi^2}{V}.
\end{align}
Since $\Lambda_{\rm IR}$ is small, the leading-order expansion of
Eq.~\eqref{eq:analytic01} yields
\begin{align}
    \int^{\Lambda_{\rm IR}}
    \frac{d^3 k}{(2\pi)^3}\,\tilde{G}_c(\vec{k})
    \approx \frac{1}{\beta m^2}\frac{1}{V},
\end{align}
hence
\begin{align}
    R_{\rm GL}^{(2)}(\Lambda_{\rm IR})
    \to
    \frac{\tilde{G}_c(\vec{k}\to 0)}{V\,\bar{\phi}^2}.
\end{align}
This matches the prescription of Eq.~\eqref{eq:R1} and naturally explains the
appearance of the $1/V$ factor.

This exercise teaches an important lesson: the two definitions of the GL ratio
agree once they include the same physical range of fluctuations. Moreover, the
regulated definition in Eq.~\eqref{eq:R2} is often superior because it allows
the IR fluctuation window to be tuned continuously.

Two remarks are in order. 
First, in textbook discussions the volume $V_\phi$ in Eq.~\eqref{eq:R1} is often
identified with the correlation volume $V_\xi = \xi^3$, with the correlation
length estimated from the curvature of the MF potential,
\begin{align}
    \xi^2 \leftrightarrow U_{\rm MF}'' .
\end{align}
Away from the actual critical region this replacement is not valid. In
particular, during a crossover the identification $V_\phi = V_\xi$ defeats the
purpose of diagnosing the breakdown of MF theory.

Second, one might attempt to define a renormalized fluctuation by subtracting
the UV contribution:
\begin{align}
    \begin{split}
        \langle \phi_R^2 \rangle_c
        &= \int^{\Lambda} \frac{d^3 k}{(2\pi)^3} \,
        \beta^{-1}
        \left( \frac{1}{\vec{k}^2 + m^2} - \frac{1}{\vec{k}^2} \right)
        \\
        &\to - \frac{m}{4\pi\beta},
    \end{split}
\end{align}
where the limit $\Lambda \to \infty$ is finite. This procedure is equivalent to
subtracting the correlator at $\xi = \infty \leftrightarrow m = 0$ 
suggested in Ref.~\cite{Friman:2011zz}. In coordinate space this scheme
corresponds to subtracting the UV singularity:
\begin{align}
    \begin{split}
        \langle \phi_R^2 \rangle_c
        &= G(\vec{x}\to 0) - G_{\rm bare}(\vec{x}\to 0) \\
        &= \beta^{-1}\lim_{r\to 0}
        \left( \frac{e^{-mr}}{4\pi r} - \frac{1}{4\pi r} \right)
        \\
        &\to -\frac{m}{4\pi\beta}.
    \end{split}
\end{align}
Although this subtraction does remove the UV dependence, it produces the unphysical
result that the square of an operator becomes negative. This is purely an
artifact of the choice of a renormalization scheme—a finite correction ultimately linked to the choice of subtraction point.

A closely analogous issue arises in the renormalization of the Polyakov loop
\cite{Nadkarni:1986cz,Kaczmarek:2002mc,Kaczmarek:2005ui}. Expanding the loop,
\begin{align}
    \begin{split}
        \ell(x)
        &= \frac{1}{N_c}\,\mathrm{tr}\, e^{i\beta A_4(x)} \\
        &\approx 1 - \beta^2 \, \frac{1}{2N_c}\mathrm{tr}\,A_4(x)^2 \\
        &\approx \exp\!\left(
        -\frac{\beta^2}{2N_c}\mathrm{tr}\,A_4(x)^2
        \right),
    \end{split}
\end{align}
one finds that certain renormalization schemes imply
\begin{align}
    \langle A_4^R(x)^2 \rangle < 0,
\end{align}
from which one would conclude that the renormalized Polyakov loop can exceed
unity—contradicting the group-theoretic constraint restricting it to the target
space parametrized by the Cartan angles~\cite{Lo:2021qkw}, e.g. for $SU(3)$:
\begin{align}
    \ell = \frac{1}{3}\left(
        e^{i\gamma_1} + e^{i\gamma_2} + e^{-i(\gamma_1+\gamma_2)}
    \right).
\end{align}
This remains a major conceptual issue in the theoretical treatment of the
Polyakov loop \cite{Lo:2013hla}. 
Although the present discussion points toward a possible resolution, we set this
issue aside and return to our main theme of developing methods that go beyond
the standard MF treatment. In the next section we examine how to incorporate the
kinetic term.

\section{Kinetic Term}

We begin with a brief review of the GL
formulation~\cite{Goldenfeld:1992qy,Cardy:1996xt,Binder_1987}. The
partition function is written as a functional integral
\begin{align}
    Z_{\rm GL} = \int D\phi_\Lambda \, e^{- \beta F_{\rm GL}[\phi]}.
\end{align}
The relevant points are as follows:
First, $F_{\rm GL}[\phi]$ is the GL functional. We use $F$ rather than $H$ to
emphasize that we are modeling a free-energy functional; the entropy
contribution is absorbed into its coefficients.

Second, $\phi_\Lambda$ indicates that field configurations are restricted to be
smooth up to a coarse-graining scale $\Lambda$, i.e.\ fluctuations with
$p > \Lambda$ (or $L < \Lambda^{-1}$) have been integrated out. Schematically,
\begin{align}
    \begin{split}
        \phi_\Lambda(\vec{x})
        &= \frac{1}{V} \sum_{\vec{x}\in \Lambda^{-3}} \phi(x)
        \\
        &\to
        \int^\Lambda \frac{d^3k}{(2\pi)^3}\,
        \tilde{\phi}(\vec{k}) e^{i\vec{k}\cdot\vec{x}} .
    \end{split}
\end{align}

Third, $Z_{\rm GL}$ is a field-theoretic partition function: all compatible
configurations are included, and the classical configuration satisfies
\begin{align}
    \frac{\delta F_{\rm GL}}{\delta \phi} = 0.
\end{align}

Our aim here is to develop
a framework that keeps the discussion as simple as Landau theory but includes
the kinetic term explicitly. For this purpose we consider a classical model
with a separable interaction kernel,

\begin{widetext}
\begin{align}
    F_{\rm GL}[\phi]
    = \int_{\vec{x}} \frac{1}{2} (\nabla \phi)^2
    + \frac{\lambda_2}{2}
         \int_{\vec{x},\vec{y}} V(\vec{x},\vec{y})\,
         \phi(\vec{x}) \phi(\vec{y})
    + \frac{\lambda_4}{4}
         \left( \int_{\vec{x},\vec{y}} V(\vec{x},\vec{y})
         \phi(\vec{x}) \phi(\vec{y}) \right)^2 .
    \label{eq:GLmodel1}
\end{align}
The equation of motion follows from

\begin{align}
    \frac{\delta F_{\rm GL}}{\delta \phi(\vec{x})} = 0 = -\nabla^2 \phi(\vec{x}) + \lambda_2 \int_{\vec{y}} V(\vec{x},\vec{y}) \phi(\vec{y}) 
    + \lambda_4 \int_{\vec{y}} V(\vec{x},\vec{y}) \phi(\vec{y}) \int_{\vec{x}',\vec{y}'} V(\vec{x}',\vec{y}') \phi(\vec{x}') \phi(\vec{y}') .
    \label{eq:eom}
\end{align}
\end{widetext}

For a local kernel the problem reduces to the usual Landau theory. In general
the classical configuration can be obtained numerically. To illustrate the
structure analytically, and to contrast with the local case, we consider a
separable kernel,
\begin{align}
    V(\vec{x},\vec{y}) \to \bar{V}\, u(\vec{x})\,u(\vec{y}).
\end{align}
The equation of motion becomes
\begin{align}
    -\nabla^2 \phi(\vec{x})
    + (\lambda_2 \bar{V} c + \lambda_4 \bar{V}^2 c^3)\, u(\vec{x})
    = 0 ,
\end{align}
where
\begin{align}
    c = \int_{\vec{x}} u(\vec{x}) \phi(\vec{x}),
\end{align}
and we define
\begin{align}
    Q = \lambda_2 \bar{V} c + \lambda_4 \bar{V}^2 c^3 .
    \label{eq:Qcond}
\end{align}
Thus,
\begin{align}
    \nabla^2 \phi(\vec{x}) = Q\, u(\vec{x}).
    \label{eq:Qreplace}
\end{align}

For definiteness we choose
\begin{align}
    u(\vec{x}) \to u(r) = e^{-m_1 r},
\end{align}
where $m_1$ is a characteristic interaction scale. The solution is
\begin{align}
    \begin{split}
        \phi(\vec{x})
        &= \frac{Q}{m_1^2}\, f(m_1 r),
        \\
        f(x)
        &= \frac{e^{-x}(2 + x)}{x}.
    \end{split}
    \label{eq:GL01}
\end{align}
The functional $c$ becomes
\begin{align}
    \begin{split}
        c
        &= \int_0^\infty dr\, 4\pi r^2 u(r)\phi(r)
        \\
        &= 3\pi \frac{Q}{m_1^5}.
    \end{split}
    \label{eq:GL02}
\end{align}
The self-consistency condition then fixes $Q$ as
\begin{align}
    Q^2
    = \frac{
        1 - \lambda_2 \bar{V} \frac{3\pi}{m_1^5}
      }{
        \lambda_4 \bar{V}^2 \left( \frac{3\pi}{m_1^5} \right)^3
      }.
    \label{eq:GL03}
\end{align}

\begin{table}[h]
\centering
\caption{Energy dimensions of model quantities}
\label{tab1}
\begin{tabular}{|l|c|}
\hline
\textbf{model ingredient} & \textbf{energy dimension} \\ \hline
$F_{\rm GL}$            & $E$   \\
$\phi$                  & $E$   \\
$\lambda_2$             & $E^2$ \\
$\lambda_4$             & $E^3$ \\
$V(\vec{x}, \vec{y})$   & $E^3$ \\ \hline
\end{tabular}
\end{table}

This analytic model demonstrates how interactions generate a spatial profile for the classical field, encoded here in the function $f$. 
This extends the Landau picture beyond its usual focus on the purely algebraic interplay between $\lambda_2$ and $\lambda_4$. 
Because homogeneous configurations neither solve the equations of motion nor
minimize $F_{\rm GL}$, the non-trivial profile in Eq.~\eqref{eq:GL01} is
mandatory.
The variational problem can be reformulated as an effective potential
in the $c$-space: using Eqs.~\eqref{eq:Qreplace}--\eqref{eq:GL02}, the functional
reduces to a potential

\begin{align}
    F_{\rm GL}[\phi] \to F_{\rm MF}(c)
    &= -\frac{1}{2} \frac{m_1^5}{3\pi} c^2
       + \frac{1}{2} \lambda_2 \bar{V} c^2
       + \frac{1}{4} \lambda_4 \bar{V}^2 c^4,
    \label{eq:GL04}
\end{align}
where the first term arises from the kinetic contribution
\begin{align}
    \begin{split}
    K &= -\frac{1}{2} \int_{\vec{x}} \phi(\vec{x}) \, \nabla^2 \phi(\vec{x}) \\
       &\to -\int d^3 x \frac{1}{2} \phi(\vec{x}) Q u(x) = -\frac{1}{2} Q c \propto -c^2.
    \end{split}
    \label{eq:GL05}
\end{align}
This term is negative and thus lowers the free energy. We define the kinetic
energy contribution via the first line of Eq.~\eqref{eq:GL05}, namely
$-\frac{1}{2}\int_{\vec{x}} \phi\,\nabla^2\phi$. Although the surface term at
$r\to\infty$ vanishes, the gradient energy $\frac{1}{2}\int_{\vec{x}}
(\nabla\phi)^2$ is UV divergent (the integrand behaves as $ (\nabla \phi)^2 \sim 1/r^4 $ near the origin), and the divergence
theorem therefore cannot be applied. The on-shell form
$-\frac{1}{2}\int_{\vec{x}}\phi\,\nabla^2\phi$, evaluated with direct
substitution of the equation of motion $\nabla^2\phi = u$, is UV finite by
construction and captures the correct infrared contribution to the effective
potential. Its negative sign presents an interesting way to evade the
standard argument favoring homogeneity. Finally, minimizing
Eq.~\eqref{eq:GL04} reproduces Eq.~\eqref{eq:GL03}.

It is well known that static MF descriptions fail near phase transitions, 
where fluctuations and dynamical critical behavior dominate~\cite{Hohenberg,Son_Stephanov}.
Even so, many QCD phase‑diagram studies continue to employ spatially homogeneous
MF models~\footnote{In astrophysical contexts, even a consistent MF model
for a first‑order QCD transition is still lacking, and some models previously
claimed to describe deconfinement have been shown to be
inconsistent~\cite{Shukla:2025egs}.}, which artificially produce discontinuous jumps in conserved
densities—although real first‑order interfaces are smooth and governed by
gradient energy.
When spatial dependence is included, it is usually limited to soliton‑type
models~\cite{Gross:1980br,Escobar-Ruiz:2016aqv,Shu:2016rrr}. 

Our construction demonstrates that gradient terms arise naturally once interactions have finite range, making spatial structure an intrinsic part of the effective theory rather than an optional correction.
The implications for finite‑volume QCD and realistic system sizes—highlighted in
finite‑size scaling and endpoint studies~\cite{Braun:2011iz,Fraga:FS,Kovacs:2023kbv}—remain to be clarified.
Certain shapes of an inhomogeneous background field can weaken first order phase
transition to a crossover~\cite{Li:2021qcb}.
Phenomenological work on QCD droplets and nucleation has long recognized finite‑size constraints~\cite{Fraga:2003mu}, but droplet profiles are rarely solved dynamically in space.

\section{Evolution of Fixed Points}

As a final topic we examine how RG flow equations are modified when an
interaction introduces an inherent momentum scale. For concreteness we start
from a conventional three-dimensional $\phi^4$ model:

\begin{align}
    F_{\rm GL}[\phi]
    = \int_{\vec{x}} \frac{1}{2} (\nabla \phi)^2
    + \frac{ \lambda_2}{2} \int_{\vec{x}} \phi^2
    + \frac{\lambda_4}{4!} \int_{\vec{x}} \phi^4.
    \label{eq:flow01}
\end{align}
The RG flow tracks how the couplings $(\lambda_2,\lambda_4)$ evolve as
modes in the momentum shell $\Lambda \to \Lambda/b$ are integrated out,
with $b = e^{t}$ and $t \to 0^{+}$. The derivation follows standard textbook
treatments~\cite{Goldenfeld:1992qy,Cardy:1996xt,Altland_Simons_2023}; 
here we compute the one-loop momentum-shell calculation directly in $d=3$, 
rather than using a $4-\epsilon$ expansion with $\epsilon = 1$. The resulting flow equations are

\begin{align}
    \begin{split}
        \frac{d}{dt}\,\bar{\lambda}_2
        &= 2 \bar{\lambda}_2
           + \frac{\bar{\lambda}_4}{4\pi^2}\,
             \bar{\beta}^{-1}\frac{1}{1+\bar{\lambda}_2}, \\
        \frac{d}{dt}\,\bar{\lambda}_4
        &= \bar{\lambda}_4
           - \frac{3}{4\pi^2}\,
             \bar{\lambda}_4^2\,
             \bar{\beta}^{-1}\frac{1}{(1+\bar{\lambda}_2)^2},
    \end{split}
    \label{eq:flow01}
\end{align}
where the dimensionless parameters are
\begin{align}
    \bar{\lambda}_2 = \lambda_2 / \Lambda^2,
    \qquad
    \bar{\lambda}_4 = \lambda_4,
    \qquad
    \bar{\beta} = \beta \Lambda.
\end{align}
The first term in each equation is the canonical scaling from the kinetic term,
while the second originates from interactions.

Fixed points satisfy
\begin{align}
    \frac{d}{dt}\,\vec{\bar{\lambda}} = \vec{0},
    \qquad
    \vec{\bar{\lambda}}
    = \begin{pmatrix}\bar{\lambda}_2 \\ \bar{\lambda}_4\end{pmatrix}.
\end{align}
There are two solutions: the Gaussian fixed point (GFP),
\begin{align}
    \vec{\bar{\lambda}}_{\rm GFP}
    = \begin{pmatrix}0 \\ 0\end{pmatrix},
\end{align}
and the Wilson–Fisher fixed point (WFFP),
\begin{align}
    \vec{\bar{\lambda}}_{\rm WFFP}
    = \begin{pmatrix}
        -1/7 \\
        \frac{48\pi^2}{49}\,\bar{\beta}
      \end{pmatrix}.
    \label{eq:wffp01}
\end{align}
To characterize the local flow we evaluate the stability matrix,
\begin{align}
    \hat{W}
    = \frac{\partial}{\partial\vec{\bar{\lambda}}}
      \left(
      \frac{d\vec{\bar{\lambda}}}{dt}
      \right),
\end{align}
whose eigenvalues dictate the local growth (or decay) rate of an infinitesimal
perturbation. At the GFP,
\begin{align}
\hat{W}_{\rm GFP}
=
\begin{pmatrix}
        2 & -\frac{1}{4\pi^2}\bar{\beta}^{-1} \\
        0 & 1
\end{pmatrix},
\end{align}
with eigenvalues
\begin{align}
    v_{\rm GFP} = [2, 1].
\end{align}
At the WFFP,
\begin{align}
\hat{W}_{\rm WFFP}
=
\begin{pmatrix}
        5/3 &
        \frac{7}{24\pi^2}\bar{\beta}^{-1} \\
        \frac{16\pi^2}{7}\bar{\beta} &
        -1
\end{pmatrix},
\label{eq:wffp02}
\end{align}
with eigenvalues
\begin{align}
    v_{\rm WFFP} = \frac{1}{3}(1 \pm \sqrt{22}).
    \label{eq:wffp03}
\end{align}
Positive (negative) eigenvalues correspond to repulsive (attractive)
directions. The corresponding operators are relevant (irrelevant).

\subsection*{Separable Interaction}

We now examine how a separable interaction modifies the RG flows. In momentum
space we introduce the form factor
\begin{align}
    u(\vec{p}) = e^{-p^2/m_1^2},
\end{align}
and replace only the quartic vertex by
\begin{align}
   F_4 \;\to\;
   \frac{1}{4!}\lambda_4
   \int \prod_{i=1}^4 \frac{d^3 p_i}{(2\pi)^3}\,
        u(\vec p_i)\,\tilde{\phi}(\vec p_i)\,
        (2\pi)^3\delta\!\left(\sum_{i=1}^4 \vec p_i\right).
\end{align}
This choice keeps the bookkeeping of internal and external lines in Feynman
diagrams straightforward. Although the change looks minimal in momentum space,
it corresponds in configuration space to a smeared two-body interaction rather
than a strictly local contact term.

A standard motivation for introducing a form factor comes from nuclear physics — notably the Yamaguchi-type separable interaction~\cite{yamaguchi:pr1954aa} — 
where it encodes the internal structure of nuclei. 
Within a QFT framework, such form factors can be understood as capturing vertex dressing when solving Dyson equations for the 2- and 4-point functions. 
Here, however, our intent is purely practical: the form factor serves as a device to introduce nonlocality and an external scale into the RG flow. 
Note that $m_1$ is treated as an external scale which does not run. 

Because only the momentum shell contributes to the one-loop integrals, the
flow equations acquire factors of
\begin{align}
    u_\Lambda = e^{-\Lambda^2/m_1^2}.
\end{align}
The modified flows are
\begin{align}
    \begin{split}
        \frac{d}{dt}\,\bar{\lambda}_2
        &= 2\bar{\lambda}_2
           + \frac{\bar{\lambda}_4}{4\pi^2}\,
             \bar{\beta}^{-1}\frac{u_\Lambda^2}{1+\bar{\lambda}_2}, \\
        \frac{d}{dt}\,\bar{\lambda}_4
        &= \bar{\lambda}_4
           - \frac{3}{4\pi^2}\,
             \bar{\lambda}_4^2\,\bar{\beta}^{-1}
             \frac{u_\Lambda^4}{(1+\bar{\lambda}_2)^2}.
    \end{split}
\end{align}
The GFP remains at $\vec{\bar{\lambda}}_{\rm GFP} = \begin{pmatrix}0 \\
0\end{pmatrix}$. The WFFP moves to
\begin{align}
    \vec{\bar{\lambda}}_{\rm WFFP}
    =
    \begin{pmatrix}
      -\frac{1}{1+6 u_\Lambda^2} \\
      48\pi^2\bar{\beta}
      \left(
          \frac{1}{1+6u_\Lambda^2}
      \right)^2
    \end{pmatrix},
\end{align}
which reduces to Eq.~\eqref{eq:wffp01} as $u_\Lambda \to 1$. 
In the opposite limit $u_\Lambda \to 0$, the WFFP approaches the
asymptotic value $[ -1, 48\pi^2 \bar{\beta} ]$, which (as we will see from the stability matrix)
leads to an unusual flow pattern.

The stability matrix evaluated at WFFP is given by 
\begin{align}
    \hat{W}_{\rm WFFP} = 
    \begin{pmatrix}
        2-\frac{1}{3u_\Lambda^2} & \frac{1}{24\pi^2}\bar{\beta}^{-1}
        (1+6u_\Lambda^2)  \\
        \frac{16\pi^2 \bar{\beta}}{1+6u_\Lambda^2} u_\Lambda^{-2} & -1
    \end{pmatrix},
\end{align}
and the eigenvalues can be computed:
\begin{align}
    v_{\rm WFFP} = \frac{1}{6 u_\Lambda^2} \left( 3 u_\Lambda^2 - 1 \pm \sqrt{81
    u_\Lambda^4 + 6 u_\Lambda^2 + 1}\right).
    \label{eq:wffp04}
\end{align}
These eigenvalues vary continuously as $u_\Lambda$ is dialed.

\begin{figure}
    \includegraphics[width=0.5\textwidth]{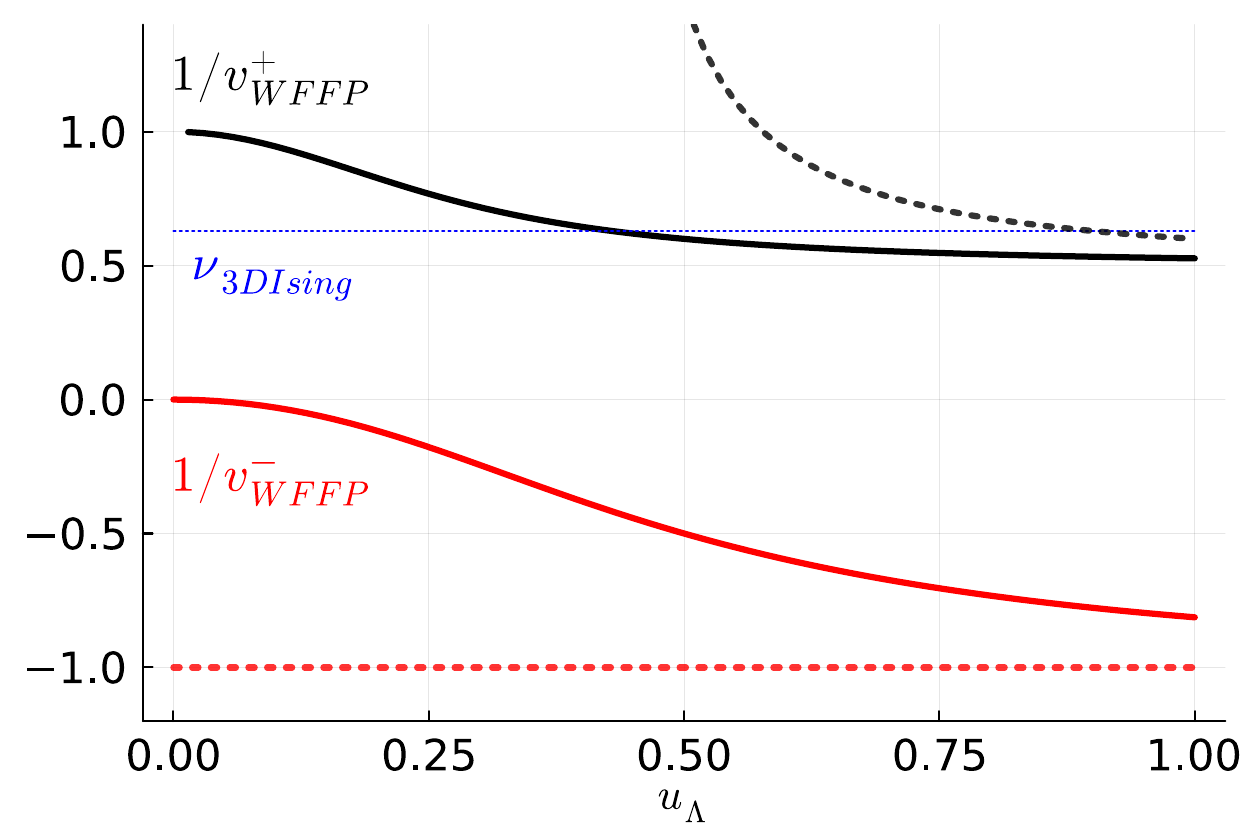}
    \caption{Inverse eigenvalues of the stability matrix at the WFFP as a
    function of shell form factor $u_\Lambda$. 
    Solid lines correspond to the full flow in Eq.~\eqref{eq:flow01}; 
    dashed lines to the truncated flow in Eq.~\eqref{eq:flow02}.}
    \label{fig:fig2}
\end{figure}

In the theory of critical phenomena, the correlation-length exponent $\nu$ is
set by the positive eigenvalue of the stability matrix via~\cite{Cardy:1996xt}
\begin{align}
    \nu = \frac{1}{v^{(+)}_{\rm WFFP}}.
\end{align}
Subleading corrections to scaling enter with exponent
\begin{align}
    \frac{ \vert v^{(-)}_{\rm WFFP} \vert}{v^{(+)}_{\rm WFFP}}.
\end{align}
The resulting exponents are shown in Fig.~\ref{fig:fig2}.

The form factor introduces some rich features. 
One of those is a tunable effective critical exponent, starting from a small
(positive) correction to the mean field result $\nu_{\rm MF} \approx 0.5$ at
$u_\Lambda = 1$ towards $\nu \approx 1$ as $u_\Lambda \to 0$. 
For reference we also show $\nu_{\rm 3D Ising} \approx 0.63$. 

At the same time, the negative eigenvalue grows without bound in magnitude as
$u_\Lambda$ decreases, signalling an increasingly fast attraction along
the irrelevant direction. Inspecting the eigenvector shows that this fast mode
is dominated by the $\bar{\lambda}_4$ direction, so trajectories are rapidly slaved to a
limiting value of $\bar{\lambda}_4 \to 48\pi^2\bar{\beta}$.

What is special here is that the remaining eigenvalue stays finite
($v_{\rm WFFP}^{(+)} \to 1$), yet its eigenvector also points predominantly along
the $\bar{\lambda}_4$ direction. The linearized flow therefore becomes strongly
non-orthogonal: after a fast collapse, the subsequent
slow drift remains almost tangential to the same $\bar{\lambda}_4$ axis.

To further probe the scheme dependence of the stability analysis, we also
include a common truncation in which only the leading $\bar{\lambda}_2$
correction is retained. In this approximation the flow equations become
\begin{align}
    \begin{split}
        \frac{d}{dt}\,\bar{\lambda}_2
        &= 2\bar{\lambda}_2
           + \frac{\bar{\lambda}_4}{4\pi^2}\,
             \bar{\beta}^{-1} u_\Lambda^2\,(1-\bar{\lambda}_2), \\
        \frac{d}{dt}\,\bar{\lambda}_4
        &= \bar{\lambda}_4
           - \frac{3}{4\pi^2}\,
             \bar{\lambda}_4^2\,\bar{\beta}^{-1} u_\Lambda^4 .
    \end{split}
    \label{eq:flow02}
\end{align}
The corresponding critical exponents are shown as dashed lines in
Fig.~\ref{fig:fig2}. Interestingly, within this truncated scheme the value at
$u_\Lambda=1$ starts at $3/5\simeq 0.6$, i.e.\ closer to the 3D Ising benchmark
than the full scheme. This naturally raises the question in what sense the full
scheme should be viewed as an improvement.

\begin{figure}
    \includegraphics[width=0.5\textwidth]{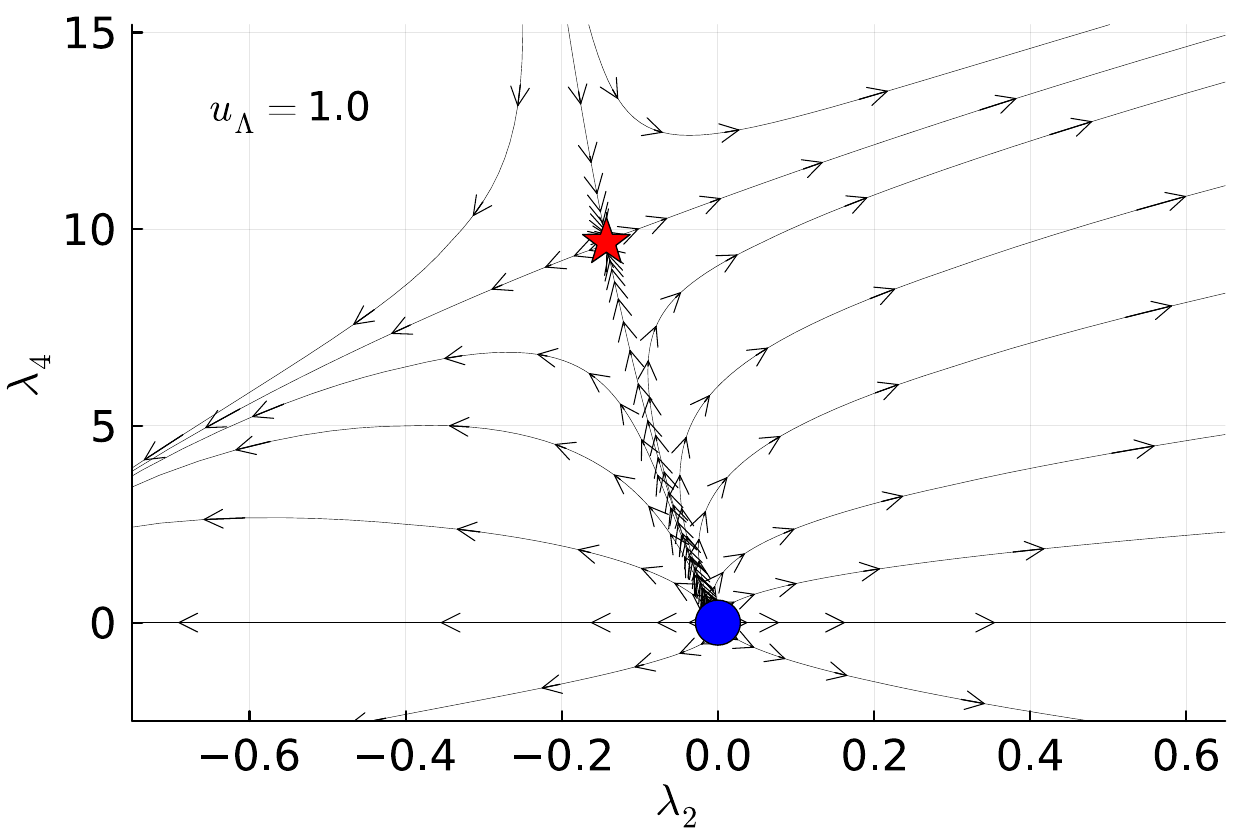}
    \caption{RG flow at $u_\Lambda=1$. Red star: WFFP. Blue point: GFP. The GFP is fully
repulsive, whereas the WFFP has one attractive direction.}
    \label{fig:fig3}
\end{figure}

Part of the answer is that the truncated scheme has a limited range of
validity: the critical exponent $1/\nu_+$ diverges at $u_\Lambda=1/\sqrt{6}
\simeq 0.4$, where the positive eigenvalue changes sign, causing the WFFP to
jump from $[-\infty,\,48\pi^2\bar{\beta}]$ to $[+\infty,\,48\pi^2\bar{\beta}]$.
This problematic behavior is corrected in the full scheme by retaining the
$\bar{\lambda}_2$ dependence in the loop propagators to all orders.
Moreover, the full scheme allows the negative eigenvalue to evolve and induces
strong mode mixing through the off-diagonal entries of the stability matrix ---
effects that are absent in the truncated treatment.

\begin{figure}
    \includegraphics[width=0.48\textwidth]{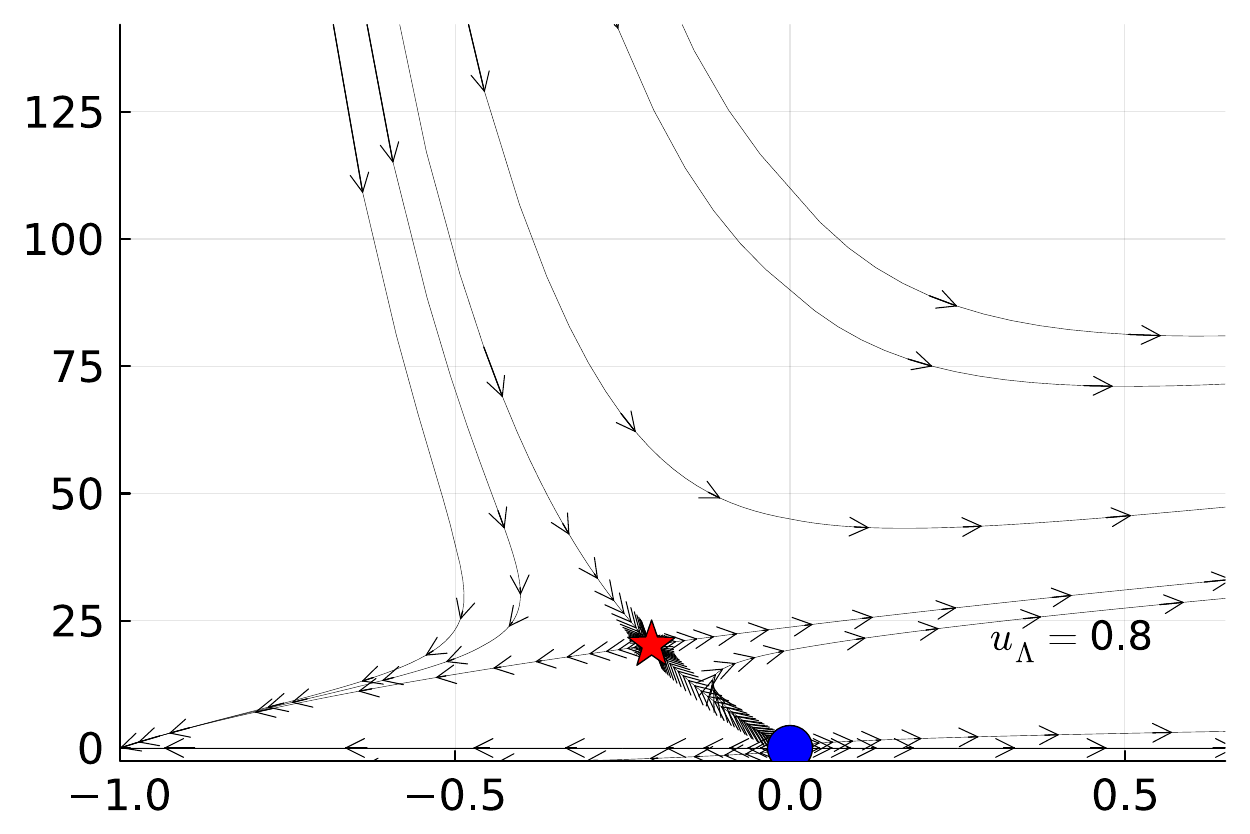}
    \includegraphics[width=0.48\textwidth]{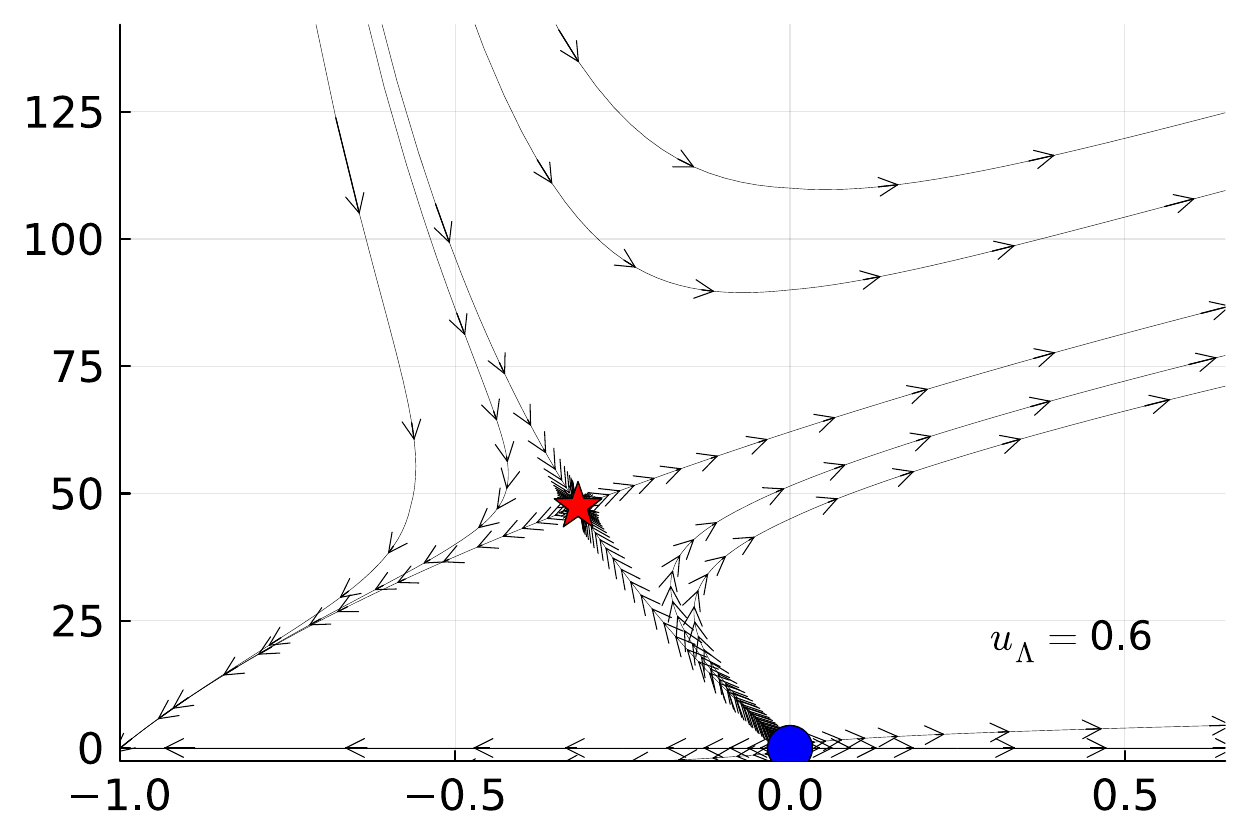}
    \includegraphics[width=0.48\textwidth]{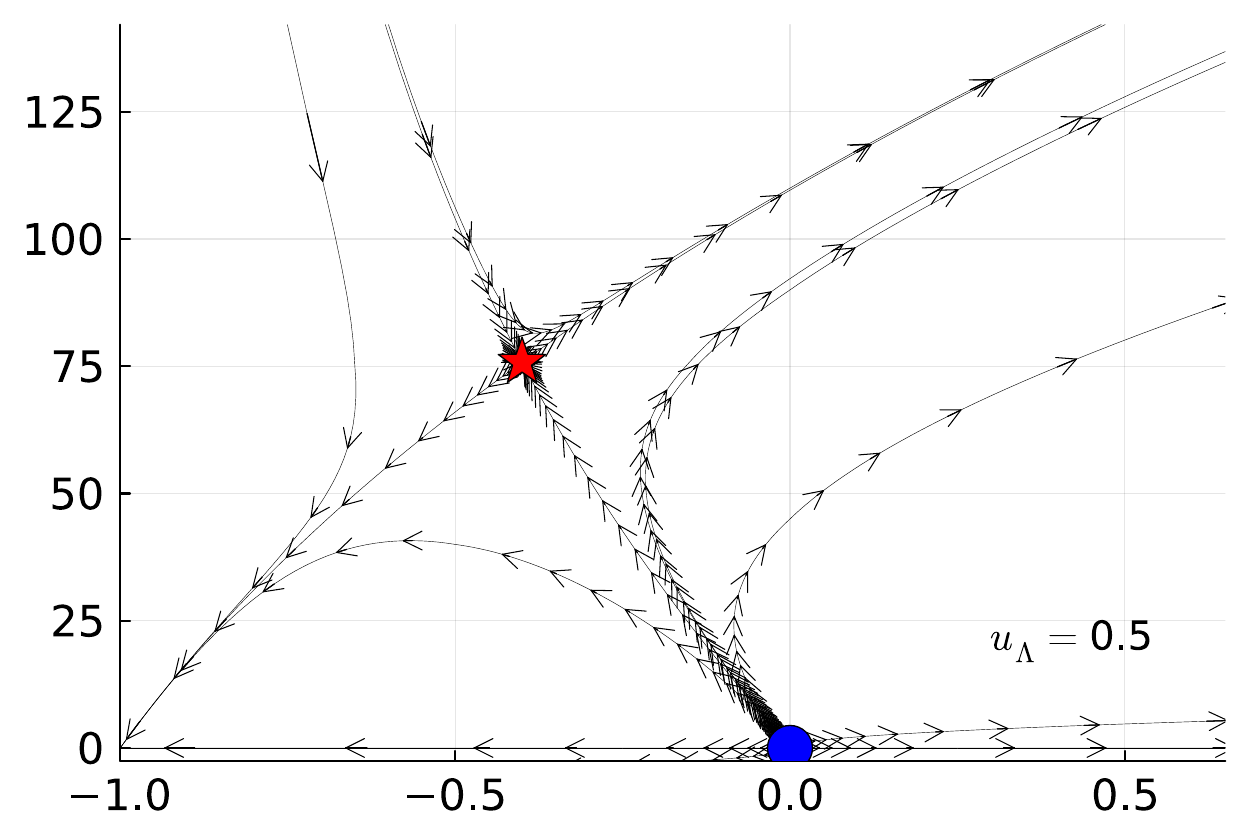}
    \includegraphics[width=0.48\textwidth]{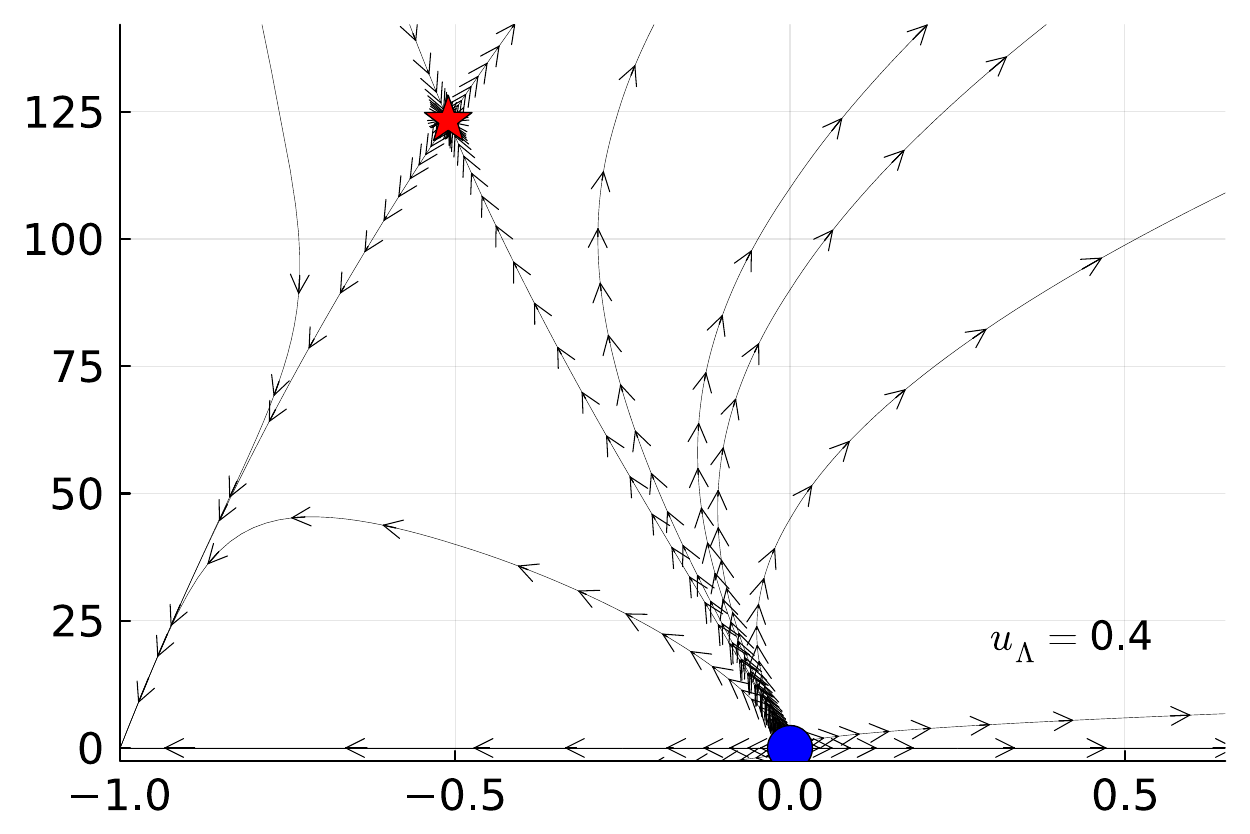}
    \caption{Evolution of the fixed points as $u_\Lambda$ decreases. The flow rotates
substantially and the WFFP moves upward, approaching the asymptotic value
$[-1,\,48\pi^2\bar{\beta}]$ in the limit $u_\Lambda\to0^+$.}
    \label{fig:fig4}
\end{figure}

Figs.~\ref{fig:fig3} and~\ref{fig:fig4} summarize the RG flow and the
evolution of the WFFP as $u_\Lambda$ is varied. The GFP (blue point) and WFFP
(red star) are marked for reference. All trajectories move away from the GFP,
since both eigenvalues of its stability matrix are positive and the GFP is
therefore fully repulsive. At the WFFP, one eigenvalue is negative, and the
flow approaches the fixed point along this stable direction. The form factor
rotates the flow pattern and shifts the location of the WFFP, which
eventually approaches the asymptotic position on the boundary
$[-1,\,48\pi^2\bar{\beta}]$. In this regime the flow becomes strongly
non-orthogonal: both eigendirections align largely along the $\bar{\lambda}_4$
axis. 

Two remarks of caution are in order. First, the drift of the WFFP toward the
propagator pole $1+\bar{\lambda}_2\simeq 0$ in the small-$u_\Lambda$ regime is
likely to be a scheme artifact rather than a genuine infrared feature. Note that the
limit $u_\Lambda\to 0^+$ is not equivalent to setting $u_\Lambda=0$: the former
refers to the asymptotic regime of the deformed flow, whereas $u_\Lambda=0$
reduces the flow to the GFP.

Second, although this family of flows admits a tunable effective critical
exponent, this should not be interpreted as a change of universality class. A
distinct universality class is usually tied to a distinct infrared
fixed-point structure --- as happens in models with genuinely long-range
kernels, where RG analyses yield long-range fixed points and a range of
critical behavior controlled by the interaction range, together with a
crossover back to short-range criticality~\cite{PhysRevLett.29.917,
RevModPhys.95.035002}. By contrast, the present rudimentary construction
deforms the flow within the same truncation without generating an additional
fixed point. 
Exploring kernels with genuinely long-range infrared structure and the
associated fixed-point scenarios is beyond the scope of these notes, but
represents a natural and potentially rich extension of the present framework.

\section{Conclusions}

We have developed several theoretical ideas that go beyond the standard MF framework. 
Rather than emphasizing universality solely through fixed-point exponents, we focus on diagnostics that are directly tied to the actual behavior of the system.

Once gradient terms or finite-range interactions are included, spatial structure
of the order parameter emerges naturally. Even minimal models generate
nontrivial spatial profiles. This shows that spatial variation is not an exotic
effect, but an intrinsic part of any consistent description beyond homogeneous
MF theory.

The present analysis demonstrates that a simple separable form factor ---
introducing nonlocality and an external scale into the flow --- is sufficient
to shift fixed points and modify the effective critical exponents. This
suggests that nonlocal interactions with genuine long-range infrared
structure merit closer examination: within a refined truncation, they may
enable a true change of universality class and the emergence of additional
nontrivial fixed points.

Ultimately, tools are only as useful as the problems they are applied to. The framework developed here sets the stage for studying interaction kernels motivated by QCD and other many‑body systems, where finite‑range effects, spatial inhomogeneities, and fluctuation windows are central ingredients. 
Such analyses can clarify which microscopic inputs control the emergence of universal behavior and where new universality structures may appear as interaction ranges or fluctuation scales are varied.

\section*{Acknowledgments}

The author thanks Song Shu and Thomas Klaehn for helpful comments, and the referee for pointing out an error in the original calculation, which has now been corrected. 
The author also acknowledges the Wroc\l{}aw Centre for Networking and Supercomputing for providing accessible computational resources, as well as partial support from the Polish National Science Center (NCN) under Opus grant no. 2022/45/B/ST2/01527.

\bibliography{ref}

\end{document}